\documentclass[useAMS,usenatbib]{mn2e}
\usepackage{graphicx}% Include figure files
\usepackage[bookmarks,colorlinks]{hyperref}
\usepackage{amsmath}

\title[Chemical abundances and spatial distribution of LGRBs]{Chemical abundances and spatial distribution of Long Gamma-Ray Bursts }
\author[M. C. Artale, L. J. Pellizza $\&$ P. B. Tissera]{M. C. Artale$^{1,2}$\thanks{E-mail:
mcartale@iafe.uba.ar} , L. J. Pellizza$^{1,2}$ and P. B. Tissera$^{1,2}$ \\
$^{1}$Instituto de Astronom\'ia y F\'isica del Espacio,C.C. 67 Suc. 28, C1428ZAA Ciudad de Buenos Aires, Argentina.\\
$^{2}$Consejo Nacional de Investigaciones Cient\'ificas y T\'ecnicas, Rivadavia 1917, C1033AAJ Ciudad de Buenos Aires, Argentina.
}
\begin{document}

%\date{Accepted 1988 December 15. Received 1988 December 14; in original form 1988 October 11}

\pagerange{\pageref{firstpage}--\pageref{lastpage}} \pubyear{2011}

\maketitle

\label{firstpage}

\begin{abstract}
 We analyse the  spatial distribution within host galaxies and chemical properties of the progenitors of  Long Gamma Ray Bursts 
as a function of
redshift. By using  hydrodynamical cosmological simulations which include star formation, Supernova feedback and
chemical enrichment and based on the hypothesis of the collapsar model with low metallicity, we investigate
the progenitors in the range $0<z<3$.
Our results suggest that the sites of these phenomena tend to be located in the central regions of the hosts
at high redshifts but move outwards for lower ones.
We find that scenarios with low metallicity cut-offs best fit current observations.
For these scenarios Long Gamma Ray Bursts tend to be [Fe/H] poor and show a strong $\alpha$-enhancement evolution towards lower values
as redshift decreases.
The variation of typical burst sites with redshift would imply that
they might be tracing different part of galaxies at different redshifts. 

\end{abstract}


\section{Introduction}

Long gamma-ray bursts \citep[LGRBs, see the reviews by][]{Mes06,Att09} are
energetic radiation events, lasting between 2 and $\sim$1000 seconds, and
with photon energies in the range of keV--MeV. 
Our current understanding of these sources indicates
that the emission is produced during the collapse of massive
stars, when the recently formed black hole accretes the debris of the stellar
core. During the accretion, highly collimated ultrarelativistic jets
consisting mainly of an expanding plasma of leptons and photons
(fireball) are launched, which drill the stellar envelope. Internal shocks in
the fireball accelerate leptons and produce the $\gamma$-ray radiation through
synchrotron and inverse Compton processes. External shocks from the
interaction of the jets with the interstellar medium produce later emission at
lower energies, from X-rays to radio (afterglow). Optical afterglow spectra
allowed the measurement of LGRB redshifts \citep{Met97}, locating these sources
at cosmological distances ($z \sim 0.01-8.2$), and revealing that their
energetics is similar to that of Supernovae (SNe). Some
LGRBs have indeed been observed to be associated to hydrogen-deficient,
type Ib/c supernovae  \citep*[e.g.][]{Gal98,Hjo03,Woo06,Starling2010}. Afterglows allowed also the
identification of LGRB host galaxies (HGs), which turned out to be mostly low-mass,
blue and subluminous galaxies with active star formation
\citep*{LeF03,Chr04,Pro04,Sav09}.

Although the general picture is clear enough, its details are still a matter
of discussion. Among other unanswered questions, the exact nature of the LGRB
stellar progenitors is still being debated. Stellar evolution models provide a
rough picture of the production of a LGRB in a massive star. According to the
{\em collapsar} model \citep*{Woo93,Mc99}, LGRBs are produced during the
collapse of single Wolf-Rayet (WR) stars. WR stars have massive cores that
may collapse into black holes, and are fast rotators, a condition needed to
support an accretion disc and launch the collimated jets. WRs have also large
mass-loss rates, needed to lose their hydrogen envelope before
collapsing, that would otherwise brake the LGRB jet. This model agrees with the
observed association between LGRBs and hydrogen deficient SNe. However, WRs
large mass-loss rates imply large angular momentum losses that would brake
their cores, which would inhibit the production of the LGRB. To overcome this
problem, \citet*{Hir05} proposed low-metallicity WRs (WOs) as progenitors. WOs
have lower mass-loss rates, diminishing the braking effect, but also preventing
the loss of the envelope. Another possibility was proposed by 
\citet*{Yoo06}. According to these authors, low-metallicity, rapidly
rotating massive stars evolve in a chemically homogeneous way, hence burning
the hydrogen envelope, instead of losing it. Low-metallicity progenitor models
are consistent with different pieces of evidence. First, the works of \citet{Mey05} and \citet{Geor09}
show that the collapse of high-metallicity stars produces mainly neutron stars,
while those of low-metallicity stars form black holes. Second, LGRB HGs have
been found to be low-metallicity systems \citep{Fynbo2003,LeF03,Chr04,Pro04,Sav09}. Finally,
the analysis of the statistical properties of the population of LGRBs suggests
that their cosmic production rate should increase with respect to the cosmic star
formation rate at high redshift, which could be explained as an effect of the low
metallicity of the progenitors, combined with the cosmic metallicity evolution
\citep*{Dai06,Sal07}.
Another possibility for WR to lose the envelope without losing too much
angular momentum is to be part of binary systems as proposed by \citet{Fry05}.

Understanding the nature of LGRB progenitors is beyond the interest of only
stellar evolution, black hole formation, and high energy astrophysics. The
visibility of LGRBs up to very high redshifts ($z > 8$), allows their use 
as tools to explore star formation and galaxy evolution in the early Universe. On the
other hand, observations of the environment and HGs of LGRBs could reveal
important clues about the progenitors of these phenomena. Given that star
formation shifts outward within a galaxy due to the depletion of gas in the central
regions as the galaxy evolves, that the interstellar medium of galaxies is
not chemically homogeneous, and that the chemical enrichment 
is affected by variations of the star formation rate and the production of different
types of SNe, it is expected that both the LGRB positions within a galaxy and
the chemical properties of the environment in which LGRBs occur depend on
redshift and on the metallicity of the LGRB progenitors.

Using high-precision astrometry, \citet*{Bloom2002}
and \citet{Blinnikov2005}
have measured the positions of $\sim$35 LGRBs with respect to the centres of
their hosts, supporting the collapsar model against the (now disproved) neutron
star merger model. The question of the metallicity dependence of LGRB
progenitors could also be investigated comparing these data with model
predictions. The chemical abundances of LGRB circumburst and HG environments were
investigated by several authors \citep*{Pro07,Niino09,Sav09,Schady2010}. However,
only in a few cases of low-redshift bursts a direct measure of the metallicity
of the star-forming region that produced the LGRB is available. At intermediate
redshift observers usually measure the mean HG metallicity, while at high
redshift they must resort to GRB-DLA techniques, which give the metallicity of
galactic clouds intercepting the line of sight to the LGRB, but not necessarily
associated with the burst itself \citep{Pro07,Rau10}.

In this paper, we use cosmological hydrodynamical simulations which include star formation and
SN feedback to investigate the predictions of
different progenitor scenarios regarding the positions of LGRBs and the
chemical abundances of their environment. Since galaxy formation is a highly
non-linear process, cosmological numerical simulations
\citep{Kat91,Nav93,Mosconi2001,Spr03,Scannapieco2005,Scannapieco2006} are the
best tools to investigate these LGRB properties. In the past, this method has
been used by several authors to investigate different aspects of the LGRB environment and HGs.
\citet*{Cou04} have shown that requiring HGs to have high star formation
efficiency, the observed HG luminosity function can be reproduced.
\citet{Nuz07} developed a Monte Carlo simulation to synthesize LGRB and HG
populations in hydrodynamical simulations of galaxy formation, in the
framework of the collapsar model. They have found that a bias to
low-metallicity progenitors ($Z < 0.3 Z_\odot$) is needed to explain the
observed properties of HGs. 
\citet{Cam09} and \citet*{Chisari2010} used semi-analytical models of
galaxy formation to study the properties of HG populations.
Particularly \citet{Chisari2010} developed a new approach to model the detectability
of the LGRBs.   
Both teams explored
models with mass and metallicity cut-offs for LGRB progenitors, finding that
models with a metallicity cut-off could explain the HG properties, and hence
supporting previous claims that LGRBs are biased tracers of star formation.
However in these semi-analytical models, the spatial distribution
of individual stellar populations within HGs cannot be investigated.
The chemical abundances of LGRB-DLAs
were investigated using numerical simulations by \citet{Pontzen2010}, finding that the
clouds producing the absorption lie at galactocentric distances of the order of
1~kpc.

In this work, we use cosmological numerical simulations of galaxy formation to
construct synthetic LGRB populations, which allow us to investigate the
properties of individual stellar populations within HGs. Our simulations
are similar to those of \citep{Nuz07}, but with a higher resolution, and
include the effects of the energy feedback of SNe into the interstellar medium.
The metallicities of each stellar populations can be estimated, and used to construct different
metallicity-dependent scenarios for LGRB production within the collapsar model.
As stated by \citet{Chisari2010}, the detectability of LGRBs and their HGs is
an important aspect that should not be disregarded for a proper comparison with
the observed samples, hence we included it in our population synthesis in the
same way as these authors.

This work is organized as follows. In Sections~\ref{sim} and \ref{mod} we
describe the cosmological simulations of galaxy formation used, and our LGRB
population synthesis models, respectively. We present our results and compare
the to available observational data in Section~\ref{res}. Finally, in
Section~\ref{con} we present our conclusions.

\section{Simulations} \label{sim}

We analyse hydrodynamical cosmological simulations  performed with a version of {\small GADGET-3} which includes
star formation, metal-dependent cooling, chemical enrichment,
multiphase gas and Supernova feedback \citep[for further details see][]{Scannapieco2005,Scannapieco2006}
The simulated regions represent periodic volumes of  10 Mpc $h^{-1}$ side  and 
are consistent with a $\Lambda$-CDM universe with the following cosmological parameters:
 $\Omega_{\Lambda}$=0.7, $\Omega_{\rm m}$=0.3,
$\Omega_{\rm b}$=0.04, $\sigma_{8}$=0.9 and 
${\rm H}_{0}$=100 $h \,\,{\rm km \,\, s^{-1} \, Mpc^{-1}}$ where $h$=0.7.

The feedback model considers Type II and Type Ia Supernovae (SNII and SNIa, respectively).
The energy per SN event released into the interstellar medium  is $0.7\times10^{51}\,{\rm erg}$.
The model assumes that stars with masses greater than $8M_{\odot}$ end their life as SNII  with lifetime $\approx 10^{6}$yr.
Lifetimes for the progenitors of SNIa are 
 randomly selected in the range  $0.1-1$ Gyr. The chemical yields for SNII are given by \citet{Woosley1995}  while
those of SNIa correspond to the W7 model of \citet{Thielemann1993}.
Initially gas particles are assumed to have primordial abundances of X$_{\rm H}$=0.76 and X$_{\rm He}$=0.24.
The chemical algorithm follows the enrichment by  
12 isotopes: $^1$H, $^2$He, $^{12}$C, $^{16}$O, $^{24}$Mg, $^{28}$Si, $^{56}$Fe, $^{14}$N, $^{20}$Ne,
$^{32}$S, $^{40}$Ca and $^{62}$Zn \citep{Mosconi2001}.
We would like to stress that this model has proven to be successfull at regulating the star formation activity
and at driving  powerful mass-loaded galactic winds without the need to introduce mass-depend parameters \citep{Scannapieco2008}.

We analyse two simulations: S230 and S320, which have been also used by \citet{deRossi2010a} to study
the Tully-Fisher relation obtaining very good agreement with observations. 
S230 has initially $2\times230 ^{3}$  with  dark matter masses of 
 $5.93\times 10^{6}\,{M}_{\odot}\,h^{-1}$ and initial 
 gas mass of  $9.1\times 10^{5}\,{M}_{\odot}\,h^{-1}$.
S320 initially has  $2\times320^{3}$  with 
 dark matter of $2.20\times 10^{6}\,{M}_{\odot}\,h^{-1}$ and  initial gas mass of 
$3.4\times 10^{5}\,{M}_{\odot}\,h^{-1}$. S320 was only run to $z\approx 2$ because of lack of computational power. 
We use this simulation to assess possible numerical resolution problems. 

From the general mass distribution, we select virialized structures by using the friends-of-friends technique and then 
identify all substructures  within the virial radius by applying the {\small SUBFIND} algorithm \citep{Springel2001}.
We select as simulated  galaxies those substructures sampled with more than 3000 particles.
  
\citet{deRossi2010} found that the mass-metallicity relation (MZR) of galaxies in these simulations differs at low redshifts from
that reported by \citet{Tremonti2004} so that galaxies have lower mean metallicity than observed 
although the shape of the observed MZR is
very well reproduced. Because of this, we renormalized  the simulated abundances to make
them consistent with observations. For that purpose, we adopted the results of \citet{Maiolino2008}
 who proposed a model to describe the evolution of observed MZR which matched available 
observations at z=0.07, 0.7 and 2.2.  With this adjustement, our simulated MZR reproduce observations at different redshifts.
For illustration purposes, we show our analysis at  four redshifts: $z=0,1,2$ and $3$.

\section{Scenarios for LGRBs} \label{mod}

To construct synthetic LGRB populations from the stellar populations described by the simulations we adopt 
the collapsar model, in which LGRB progenitors are massive stars  
possibly with low metallicity. We investigate four scenarios in which progenitors have a mass 
greater than a certain minimum $m_{\rm min}$. For scenario 1 this is the only condition, while for the others a maximum metallicity $Z_{\rm c}$
is assumed for the progenitors. The values of $m_{\rm min}$ and $Z_{\rm c}$ were taken from \citet{Chisari2010}, who derived them 
by fitting the LGRB rate observed by BATSE, and are listed in Table~\ref{scenarios2}.   

We estimate the number of massive stars in each simulated galaxy at each analysed redshift
 by assuming a Initial Mass Function given by \citet{Salpeter1955}. We included all stars born within $\tau_{\rm c}=100\,{\rm Myr}$. This time
interval is larger than the mean lifetime of SNII progenitors but it allow us to minimize numerical fluctuations and it is
small compared to SFR variations.
 The selected progenitors defined the scenario 1.  For scenarios 2, 3, and 4
we impose a requirement on the mean metallicity, considering only new born stars with $Z<Z_{\rm c}= 0.6, 0.3, 0.1$, respectively.

Following \citet{Chisari2010},
for each stellar population represented by a particle $p$ with mass $m_{*}(p,z)$ at redshift $z$ satisfying the above requirements, 
we calculated the number of LGRBs produced as the number of stars with $m>m_{\rm min}$,

\begin{equation}
 N(p,z)=m_{*}(p,z) \frac{\int_{m_{min}}^{100M_{\odot}} \xi(m) {\rm d}m}{ \int_{0.1M_{\odot}}^{100M_{\odot}} m \xi(m) {\rm d}m},
\end{equation}
\noindent where $\xi(m)$ is the Initial Mass Function with 
 $0.1 M_{\odot}$ and $100M_{\odot}$ its lower and upper mass cut-offs, respectively.

%%%%%%%%%%%%%%%%%%%%%%%%%%%%%%%%%%%%%%%%%%%%%%%%%%%%%%%%%%%%%%%%%%%%%%%%%%%%%%%%%%%%
\begin{table}
 \centering
 \caption{Properties of the four scenarios proposed for LGRBs. 
The values of minimal masses where taken from \citet{Chisari2010} with the initial mass function 
of \citet{Salpeter1955}. }
\label{scenarios2}
  \begin{tabular}{@{}llrrrrlrlr@{}}
  \hline
   Scenario        & $Z_{\rm c}$	        & $m_{\rm min}$\footnote{\citet{Chisari2010}} \\
 \hline
   Sc1             &  -		        &         $\sim$ 80${\rm M}_{\odot}$	                     \\
   Sc2             &  0.6		&         $\sim$ 54${\rm M}_{\odot}$      	                  \\
   Sc3             &  0.3		&         $\sim$ 23${\rm M}_{\odot}$              	     \\
   Sc4             &  0.1		&         $\sim$ 6${\rm M}_{\odot}$         		       \\
 \hline
\end{tabular}
\end{table}
%%%%%%%%%%%%%%%%%%%%%%%%%%%%%%%%%%%%%%%%%%%%%%%%%%%%%%%%%%%%%%%%%%%%%%%%%%%%%%%%%%%%

\noindent The intrinsic LGRB rate for a stellar population in any scenario is then
\begin{equation}
 r(p,z)=\frac{N(p,z)}{\tau_{c}}.
\end{equation}
\noindent We are interested in computing observable properties of the stellar
populations selected by LGRB observations, such as metallicites,
$\alpha$-elements abundances, and distances to their HG centre. 
As discussed by \citet{Chisari2010}, selection effects
introduced by observations can be modeled by weighting the properties of
simulated stellar populations by their contribution to the total observed LGRB
rate at the Earth. We applied the method developed by \citet{Chisari2010} to estimate the probability
that a certain LGRB  produced at a given $z$ could be observed at Earth. 
However, there might be other biases  introduced by observations which are difficult to model because of their dependence
on sensitivity and spectral bands of the detectors
and telescopes. Particularly, the afterglow observations, on which the precise positioning of LGRBs is based,
 are usually made in the optical range and could be affected by
dust absoption, biasing the samples towards low metallicity systems.
As claimed by \citet{Fynbo2009}, about 40 per cent of LGRBs might be dust obscured. 
Dust effects have not been included in our scenarios hence, caution should be taken when comparing our
results with observations. We will point out posible dust effects when appropriate.

At fixed redshift, the weights depend only on the intrinsic
LGRB rate of the corresponding stellar population, becoming

\begin{equation}
p_{\rm det}(p,z)=\frac{r(p,z)}{\sum_{p'} r(p',z)},
\end{equation}

\noindent
where the sum extends over all the stellar populations $p'$ producing LGRBs at
a given redshift. For any observable property $X(p,z)$ of these stellar
populations, its mean observed value at $z$ must then be

\begin{equation}
\langle X (z) \rangle = \sum_{p'} p_{\rm det}(p,z) X(p',z).
\label{rgalax}
\end{equation}

\section{Results and Analysis} \label{res}

\subsection{Spatial distribution of LGRBs}

We first investigate the spatial distribution of LGRBs in our scenarios.
For this purpose, we calculate the distance between the LGRB and  
the centre of mass of its galaxy $b$.
To eliminate the effects produced by the growth of galaxies as the structure in the Universe assembles,
we normalize $b$ by taking the ratio $b/r_{\rm opt}$, where $r_{\rm opt}$ is the optical radius of the galaxy, 
defined as the radius encompassing 83 per cent of its baryonic mass \citep{Tissera1998}.

\begin{figure}
\centering
\includegraphics[width=0.45\textwidth]{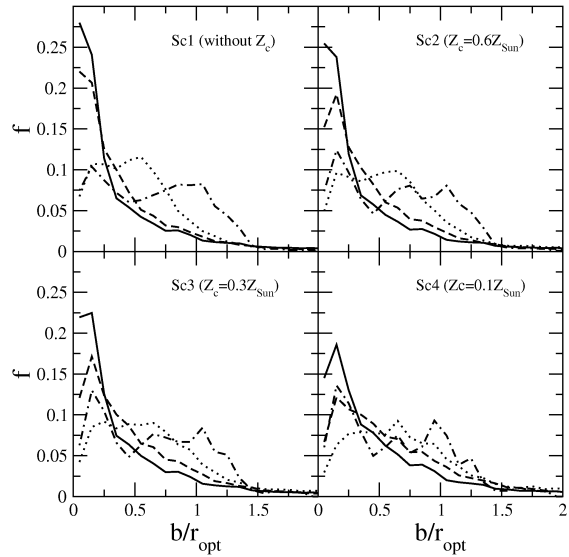}
\caption{Distribution of 
 LGRB distances to the galaxy centres normalised by the optical radius 
in our scenarios, at $z=3$ (solid lines), 2 (dashed lines), 1 (dotted line), 
and 0 (dot-dashed lines).} 
\label{bnorm}
\end{figure}

In the Fig.~\ref{bnorm} we present the distribution of $b/r_{\rm opt}$ for the LGRBs at
different redshifts ($z=0,1,2,3$), weighted by their detectability as explained in section \ref{mod}.
We observe that LGRB progenitors tend to reside in the inner regions galaxies
at high redshift, and to be progressively located at larger distances from the centre as redshift decreases. 
This is consistent with the fact that the main sites of star formation shift outwards as time evolves and
the galactic structure gets assembled in a hierarchical fashion.
This effect is stronger in our scenarios with higher $Z_{c}$, because low metallicity populations 
tend to be formed in the outer regions of galaxies which are less enriched since all simulated systems 
 exhibit metallicity gradients.

\begin{figure}
\centering
\includegraphics[width=0.45\textwidth]{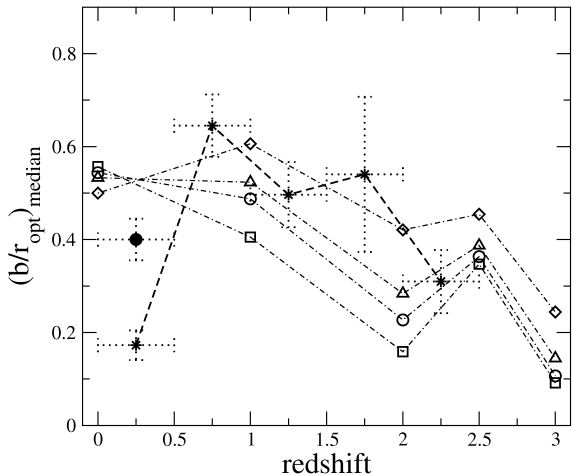}
\caption{Median LGRB distances to the galaxy centres normalised by the optical radius as a function of redshift,
in our scenarios (Sc1 squares, Sc2 circles, Sc3 triangles, Sc4 diamonds) 
compared with median calculated by the observations (stars) of \citet{Bloom2002} and \citet{Blinnikov2005}.} 
\label{bmedian}
\end{figure}

In Fig.~\ref{bmedian}, we plot the median values of $b/r_{\rm opt}$
in our scenarios as a function of redshift, together with the available observations of the LGRBs 
positions within their hosts \citep{Bloom2002,Blinnikov2005}.
These authors measure the distance of LGRB to the centre of their hosts, projected onto the plane
of the sky, and normalised by the galaxy half-light radius $r_{\rm h}$. 
As these authors point out, this normalisation is 
a crude way of deprojecting the values of $b$. To transform them into $b/r_{\rm opt}$,
we assume that LGRB hosts can be modeled by an exponential disc, for which $r_{h}=0.52r_{\rm opt}$.
Fig.~\ref{bmedian} shows that our scenarios are consistent with observations, except at very low redshifts
in which the observed median value of $b/r_{\rm opt}$ drops abruptly, while our scenarios
remain almost constant.  By analysing the LGRBs contributing to the lowest-$z$
point in Fig.~\ref{bmedian}, we find that almost half of them (3 out of 7) have
$b/r_{\rm opt}$ values consistent with zero. Interestingly, the hosts
of these LGRBs show  evidence of interaction or close companions. Hence, the presence of
nuclear star formation activity  could be explained as triggered by galaxy
interactions as suggested by observations \citep{Lambas2003} and numerical
simulations \citep{Barnes1996, Mihos1996, Tissera2000, Perez2006}. Then, the
discrepancy can  be attributed to the fact that our simulated galaxy sample does not
reflect the effects of  this mode of star formation at low redshift since  
our analysed galaxies are dominated by systems with low gas reservoir \citep{deRossi2010,derossi2011}.
A further piece of evidence for this explanation is provided by a recalculation
of the lowest-$z$ point, excluding the three quoted LGRBs (filled circle in
Fig.~\ref{bmedian}). The new point lies within $3\sigma$ of our scenarios, showing a better agreement than the original one.

The large error bars of the observations, which originate in the low number of LGRBs with precise positions,
prevent us from using a goodness-of-fit estimator to determine the scenario that better fits the observations.
However the fact that the observed values are always higher than the predictions of
scenarios Sc1 and Sc2 implies that it is very improbable that these scenarios could explain the observations.
Hence our results suggest that LGRB progenitors would have low metallicities ($Z<0.3Z_{\odot}$).
In the case that dust effects introduce important biases in the impact parameter distribution, 
the preference for low metallicity progenitors obtained from Fig.~\ref{bmedian}
would have to be re-considered. Observations providing new insights on the location of dark GRBs may help
to resolve this issue.

\subsection{Chemical abundances}

\begin{figure}
\centering
\includegraphics[width=0.45\textwidth]{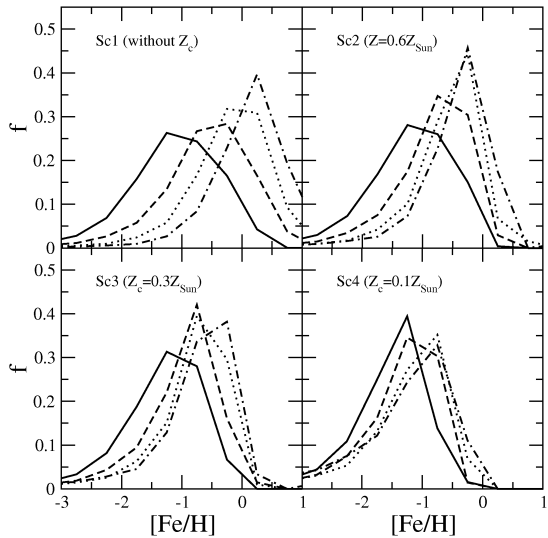}
\caption{Distribution of the relation beween Fe/H for our four scenarios at different redshifts $z=3$ (solid line), 2 (dashed line), 1 (dotted line), 
and 0 (dot-dashed line).} 
 \label{feh}
\end{figure}

To analyse the chemical abundances of the LGRB progenitors in our scenarios,
we use the ratio [Fe/H] as a measure of the iron abundance, and [Si/Fe] as a measure of 
the relative abundance of $\alpha$ elements to iron. 
In Fig.~\ref{feh} we present the distribution of [Fe/H] for the LGRB progenitors in our scenarios 
at different redshifts, wheighed by the detectability in the same way as in the previous section. 
We observe that the abundance of iron increases as redshift decreases in all scenarios.
This can be understood in terms of the chemical evolution of the interstellar medium.
As time evolves SNe contribute to the enrichment of the medium with iron, hence stellar populations born at low redshifts 
exhibit higher iron abundances.
This enrichment is stronger in scenarios Sc1 and Sc2, where the metallicity cut-offs are not so restrictive.

\begin{figure}
  \centering
  \includegraphics[width=0.45\textwidth]{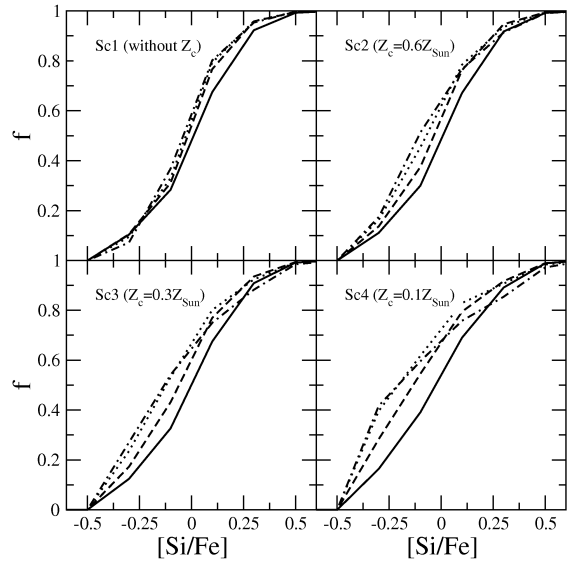}
  \caption{Cumulative histograms for [Si/Fe] for $z=3$ (solid line), 2 (dashed line), 1 (dotted line),
 and 0 (dot-dashed line) in all scenarios proposed.} 
\label{cum_sife}
\end{figure}

In Fig.~\ref{cum_sife} we present the distribution of [Si/Fe] for LGRB progenitors.
We observe that, for all our scenarios, they exibit a higher [Si/Fe]
as redshift increases, indicating an enhacement in $\alpha$ elements at high redshifts. 
This is consistent with the fact that SNIa and SNII enrich the interstellar medium.
Due to the fact that SNIa progenitors have lifetimes $\sim1{\rm Gyr}$ while those of SNII
live only $\sim10{\rm Myr}$, at high redshift only the contribution of the latter to the interstelar
medium enrichment is significant, rendering stellar populations rich in $\alpha$ elements.
As redshift decreases, the contribution of SNIa becomes important, decreasing the abundance
of $\alpha$ elements relative to iron.

\begin{figure*}
  \includegraphics[width=0.45\textwidth]{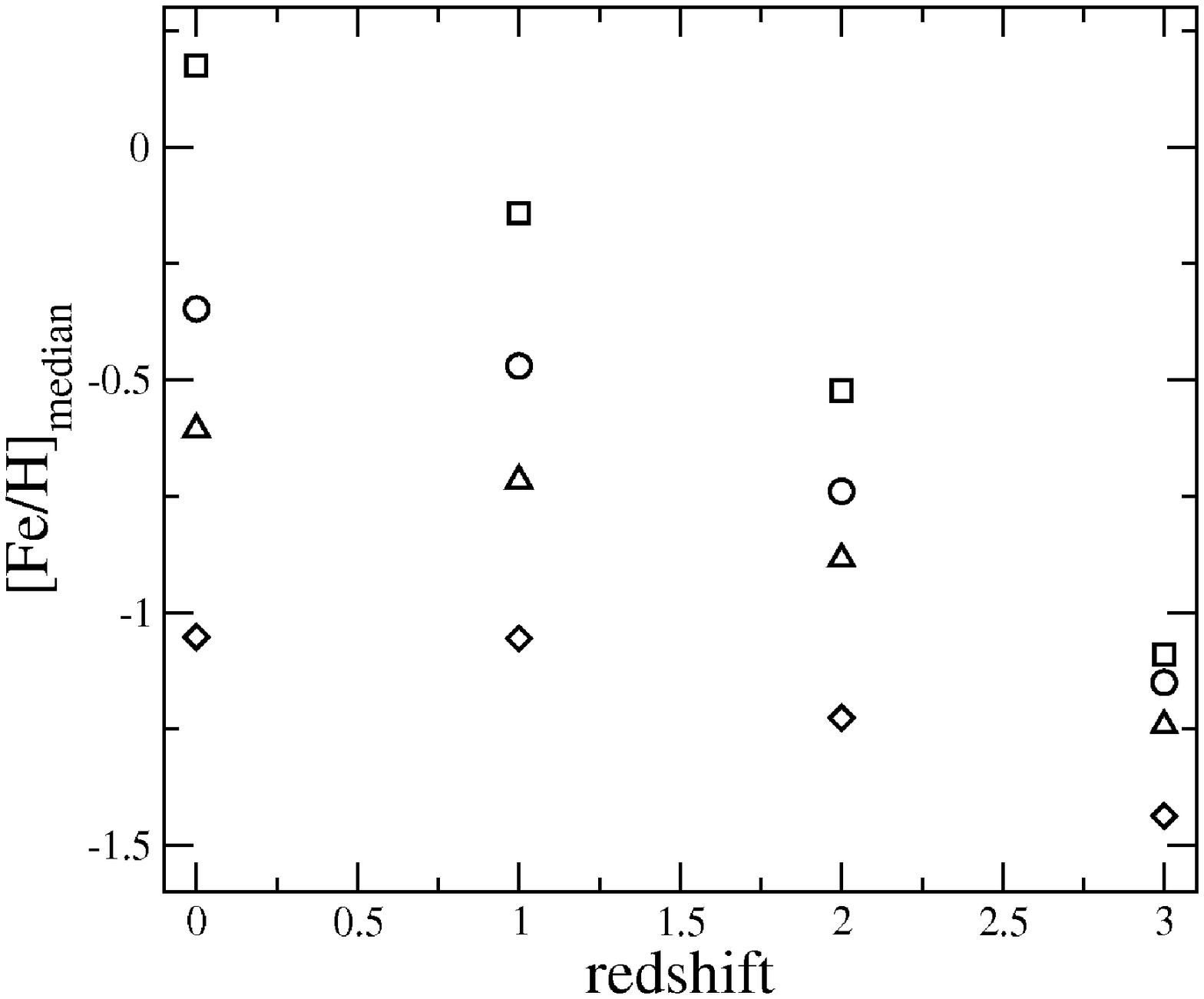}
  \includegraphics[width=0.45\textwidth]{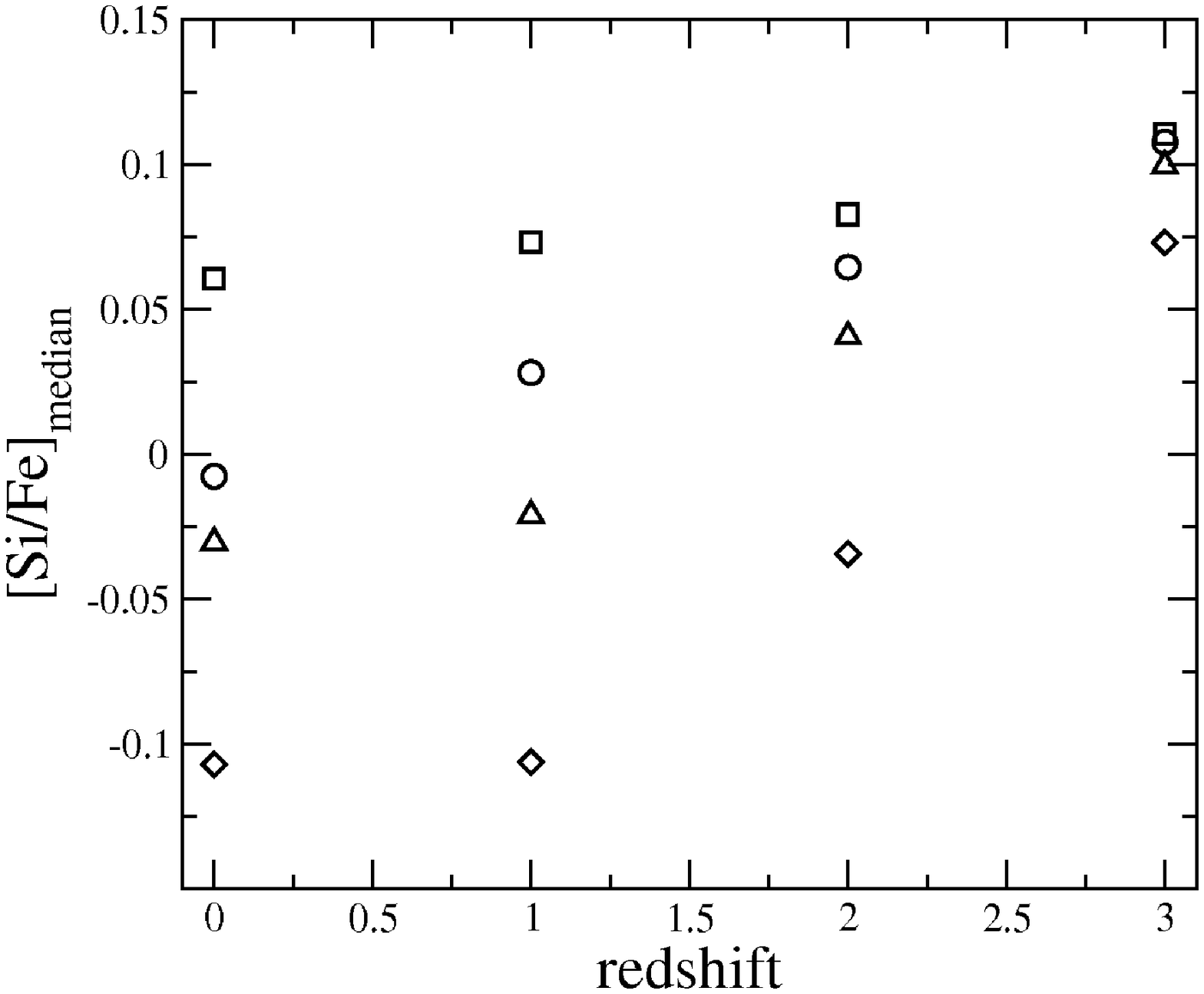}
  \caption{Evolution of median values of [Fe/H](right) and [Si/Fe](left) of LGRB progenitors in our different scenarios.  
 (Sc1 squares, Sc2 circles, Sc3 triangles, Sc4 diamonds). } 
  \label{medians_sife_feh}
\end{figure*}

The trend of [Si/Fe] to decrease towards lower redshift is also observed in Fig.\ref{medians_sife_feh} (right panel),
where we plot the median value of the ratio [Si/Fe] as a function of redshift.
In this figure we also observe that the values of [Si/Fe] of the LGRB progenitors
is lower and evolve more strongly with redshift in the scenarios where the metallicity cut-off is more restrictive. 
In the left panel of Fig.~\ref{medians_sife_feh} we show the median value of the ratio [Fe/H] as a function of redshift.
We find that the median value of [Fe/H] decreases with $Z_{\rm c}$ as expected 
for cut-offs progressively more restrictive in metallicity. 
These results indicate that the metallicity cut-off tends to eliminate old stellar populations highly enriched by SNII,
located mainly in the central regions of galaxies and originated in first outbreaks of star formation.
This interpretation agrees with the shift of the normalized impact parameters of LGRB progenitors observed 
in Fig.~\ref{bnorm}.

\section{Conclusions} \label{con}
Aiming at understanding the relation between LGRBs and star formation,
we analysed the spatial distribution and chemical abundances of stellar populations producing these phenomena.
We investigated four different scenarios for the progenitors of LGRBs based on the collapsar model with different metallicity cut-offs.

We compared the spatial distribution of LGRBs within their HGs in our scenarios and with the available observations.
We found that in all our scenarios LGRB progenitors reside on average in the outer regions of their galaxies at low redshifts,
shifting toward the centre as redshift increases. 
Scenarios favouring low metallicity progenitors tend to
produce LGRBs further out from the central regions than those allowing high metallicity progenitors.
The confrontation of our models with available observations supports scenarios with low metallicity cut-offs,
in agreement with previous results \citep{Nuz07,Cam09,Chisari2010}.
Particulary we best reproduce current available observations for a model where LGRB progenitors
are massive stars with $Z<0.3$.
Further precise LGRB position measurements would help to confirm these trends.

Regarding [Fe/H] abundances of the stellar populations producing LGRBs,
we found that in all our scenarios [Fe/H] increases as redshift decreases.
This effect is less conspicuous in the scenarios with low metallicity progenitors, 
as in these cases the metallicity cut-off restricts the chemical abundances of the stellar populations producing LGRBs.
The $\alpha$-enhancement decreases with redshift in all our scenarios, as a result of 
the different contributions of SNII and SNIa. 
Contrary to the detected trend in [Fe/H], the $\alpha$-enhancement shows a stronger evolution with redshift 
as $Z_{\rm c}$ decreases.
As previously discussed, these chemical trends can be understood within the context of chemical evolution in hierarchical 
clustering scenarios. 

Considering that the results on the spatial distribution of LGRB progenitors favours low-metallicity progenitor models,
one would expect that the iron abundance of the stellar populations producing LGRBs remains
low at all redshifts with little variations (${\rm [Fe/H]}\sim-1$). On the other hand, one 
would expect that the $\alpha$-enhancement strongly decreases with redshift (by 0.2~dex between $z=3$ and $z=0$).
This means that, if LGRBs are produced by low metallicity massive stars, their location will be shifted
on average from the central regions to the outskirts of galaxies.
If LGRBs can trace the chemical properties of the interestelar medium, they may map  
different regions of galaxies at different redshifts.
A test of these prediction could be set up as 
further dust-corrected measurements of the chemical abundances of 
the stellar populations producing LGRBs become available.

\section*{Acknowledgments}

LJP acknowledges funding by Argentine ANPCyT, through grant PICT 2006-02015 and 2007-00848.
This work was partially supported by PICT 2005-32342, PICT 2006-245 Max Planck and PIP 2009-0305.

\appendix

\end{document}